\documentclass[lettersize,journal]{IEEEtran}
\usepackage{stmaryrd}
\usepackage{amsmath}
\usepackage{amssymb}
\usepackage{psfrag}
\usepackage{graphicx}
\usepackage{epstopdf}
\usepackage{multirow}
\usepackage{booktabs}
\usepackage{siunitx}
\usepackage{cite}
\usepackage{algorithmic}
\usepackage{algorithm}
\usepackage{mathtools}
\usepackage{hyperref}
\usepackage{bm}

\usepackage{float}
\usepackage{subfigure}
\usepackage[inkscapelatex=false]{svg}
\ifCLASSOPTIONcompsoc
\else
\fi
\usepackage{comment}

\begin{document}

\title{
Learning at the Speed of Wireless: Online Real-Time Learning for AI-Enabled MIMO in NextG

}
\author{Jiarui Xu, Shashank Jere, Yifei Song, Yi-Hung Kao,  Lizhong Zheng, and Lingjia Liu
\thanks{J. Xu, S. Jere, Y. Song, Y. Kao, and L. Liu are with Wireless@Virginia Tech, ECE Department at Virginia Tech. L. Zheng is with the EECS Department at the Massachusetts Institute of Technology.
The corresponding author is L. Liu (ljliu@ieee.org).
}
}
%



\maketitle

\begin{abstract}
Integration of artificial intelligence (AI) and machine learning (ML) into the air interface has been envisioned as a key technology for next-generation (NextG) cellular networks.
At the air interface, multiple-input multiple-output (MIMO) and its variants such as multi-user MIMO (MU-MIMO) and massive/full-dimension MIMO have been key enablers across successive generations of cellular networks with evolving complexity and design challenges.
Initiating active investigation into leveraging AI/ML tools to address these challenges for MIMO becomes a critical step towards an AI-enabled NextG air interface.
At the NextG air interface, the underlying wireless environment will be extremely dynamic with operation adaptations performed on a sub-millisecond basis by MIMO operations such as MU-MIMO scheduling and rank/link adaptation.
Given the enormously large number of operation adaptation possibilities, we contend that online real-time AI/ML-based approaches constitute a promising paradigm.
To this end, we outline the inherent challenges and offer insights into the design of such online real-time AI/ML-based solutions for MIMO operations. 
An online real-time AI/ML-based method for MIMO-OFDM channel estimation is then presented, serving as a potential roadmap for developing similar techniques across various MIMO operations in NextG.

\end{abstract}


%
\IEEEpeerreviewmaketitle









\section{Introduction}





The concept of an artificial intelligence (AI)-enabled cellular network~\cite{shafin2020artificial} has attracted much attention from both academia and industry.
As a critical step, the 3rd Generation Partnership Project (3GPP) has initiated the exploration of AI in the 5G-Advanced air interface.
This trend of standardizing and deploying AI at the air interface is anticipated to continue and evolve through NextG networks. 



The growing interest in this domain mainly arises from the intrinsic issues of \emph{network complexity}, \emph{model deficit}, and \emph{algorithm deficit} as discussed in~\cite{shafin2020artificial}, but tailored towards the air interface of the NextG. 
Specifically, the air interface of the NextG (e.g., 6G and beyond) is expected to be increasingly sophisticated with complex network topologies/numerologies, non-linear device components, and high-complexity processing algorithms.
Therefore, it becomes exceedingly challenging to utilize conventional model-based approaches in a scalable and efficient manner.   
Meanwhile, AI/ML-based data-driven approaches can effectively resolve these issues, providing an appealing alternative for the design of the NextG air interface.
MIMO technology has been a prominent enabler for enhanced system performance across successive generations of cellular networks and remains a cornerstone in 5G-Advanced and NextG.
Massive MIMO techniques, particularly in the emerging context of MU-MIMO and distributed MIMO, introduce new grand challenges for conventional model-based approaches.
AI/ML can offer promising solutions to achieve attractive performance–complexity trade-offs for various MIMO operations at the air interface: attaining quality of service (QoS) requirements while maintaining acceptable inference complexity.
Recognizing the need for AI/ML integration into MIMO operations, we pose a foundational question: What types of AI/ML-enabled MIMO operations are suitable at the NextG air interface? 

\section{Types of AI-enabled MIMO operations}
The deployment of AI/ML-enabled methods for MIMO operations at the air interface must ensure reliable and low latency transception while meeting QoS requirements across a wide range of system configurations, scenarios, and possible operation adaptations.

A direct approach to meet these requirements is to employ \textbf{offline learning}. 
Specifically, offline learning refers to the process in which the AI/ML model is trained with a large amount of offline field-collected data and then deployed for real-time inference. 
Many existing works have focused on developing methods using offline learning. 
For example, deep residual learning is exploited in~\cite{LLi2020} to design an offline channel estimation approach.
However, offline-trained AI/ML models often encounter the well-known ``uncertainty in generalization'' issue~\cite{shafin2020artificial}, which can be caused by a combination of the following mismatch between offline training and online deployment: i) \emph{system configuration}, ii) \emph{scenario}, and iii) \emph{operation adaptation}.
Specifically, system configuration refers to antenna configurations, antenna numbers, and network numerologies. 
Scenarios include outdoor versus indoor, urban versus rural, and macro versus micro settings, which can change more frequently than system configurations.
In 5G NR and NextG, operations such as link/rank adaptation and scheduling are expected to be performed on a slot basis, which makes operation adaptations occur on a sub-millisecond level~\cite{std3gpp38912}.
Furthermore, the number of possible operation adaptations within each slot can be enormously large. 
For example, for a single-cell network with $18$ subbands and $10$ active users, the number of scheduling options for single-user MIMO (SU-MIMO) operation can be as large as $10^{18}$, while the same for MU-MIMO is even larger.
Therefore, it is extremely challenging if not impossible to ensure the robust generalization ability of offline learning methods across all possible cases.
Another option involves offline learning followed by online adaptation, which is a \textbf{hybrid} approach.
Specifically, this hybrid approach starts from an offline AI/ML model that is trained with substantial offline collected data and then adapted online using the limited online training data.
The online adaptation approach is developed to fine-tune the offline-trained model to different scenarios, which can mitigate poor generalization caused by the scenario mismatch.
Due to the dynamically changing operation environment (channel and/or scheduling), it is desirable for the AI/ML model updated online within the scheduling granularity utilizing over-the-air (OTA) reference signals (RS).
For instance, the scheduling granularity in 5G NR is a slot.
However, the online OTA RS (training data) within the scheduling granularity is extremely limited due to the system overhead constraint.
For cases with substantial deviation between the offline and online scenarios that induce different underlying distributions in the corresponding datasets, the AI/ML model obtained from this hybrid approach may be heavily biased toward the offline dataset leading to degraded performance during online deployment.
Recent attempts have been made at developing possible hybrid learning approaches using model-agnostic meta-learning (MAML).
The MAML algorithm aims to learn a general model initialization for fast adaptation to new tasks utilizing only a few training samples and is independent of the model architecture.
Examples of the MAML algorithm being adopted for hybrid learning include the downlink beamforming approach at the transmitter side~\cite{Yuan2021_BM_Transfer_Learning} and the channel prediction method at the receiver side~\cite{Yang2020DeepTransfer}.
These meta-learning-based methods have been effective in mitigating \emph{scenario mismatch} with a reduced amount of online training samples compared to the offline dataset.
However, the mismatch caused by the system configuration and the operation adaptation between offline and online stages, which are major mismatches, will still impede the smooth deployment of such a transfer or meta-learning-inspired hybrid approach in practice.
As an alternative, we introduce another \textbf{hybrid} approach that starts with: i) identifying an AI/ML-based approach that is completely online and real-time, i.e., relies only on the OTA RS and is updated based on their scheduling granularity, followed by ii) adding ``offline components'' to improve its overall efficiency and resiliency.
Such an online real-time AI/ML-based approach avoids the generalization issue caused by all three aforementioned mismatches due to the fact that the training and testing data share almost the same features within the scheduling granularity. 
Subsequently, as introduced in our previous work, the additional ``offline component'' can be trained using data-driven methods or directly extracted from the available domain knowledge, e.g., exploiting the pre-trained nearest neighbor binary classifier~\cite{JXu2023}, or utilizing available channel statistics~\cite{JereMILCOM2023}, to further improve the performance and learning efficiency of the online learning.
Although the hybrid learning approach that starts with a completely offline-trained model and appended with online adaptation may cater to specific MIMO operations at the air interface, in general, we posit that the alternate hybrid approach starting with a completely online and real-time trained model supplemented with offline components can mitigate generalization challenges much more effectively.
Recognizing the inherent challenge of developing a one-size-fits-all offline learning-based solution and the need to adopt a hybrid learning paradigm in general, we envision that designing a purely \textbf{online and real-time} AI/ML solution catering to specific MIMO operations is a crucial step towards a robust hybrid learning paradigm for the NextG air interface.
In this article, we focus on online real-time AI/ML-based approaches, which can serve as potential enablers in fulfilling the vision of AI/ML-enabled MIMO operations at the NextG air interface.
Specifically, this approach entails:
i) \emph{Online learning:} Training AI/ML models with only the OTA reference signals.
ii) \emph{Real-time learning:} 
Enabling AI/ML models to complete the training procedure within the scheduling granularity.
Online learning demands sample-efficient AI/ML algorithms capable of utilizing extremely limited standard-compliant RS.
Real-time learning further requires AI/ML models to have efficient training processes to meet stringent latency constraints. 
For instance, real-time learning in 5G NR refers to slot-based learning (sub-millisecond basis).

\section{Online Real-Time Learning under Channel Uncertainty}

\begin{figure*}[t] 
    \centering
   \includegraphics[width=0.8\textwidth] {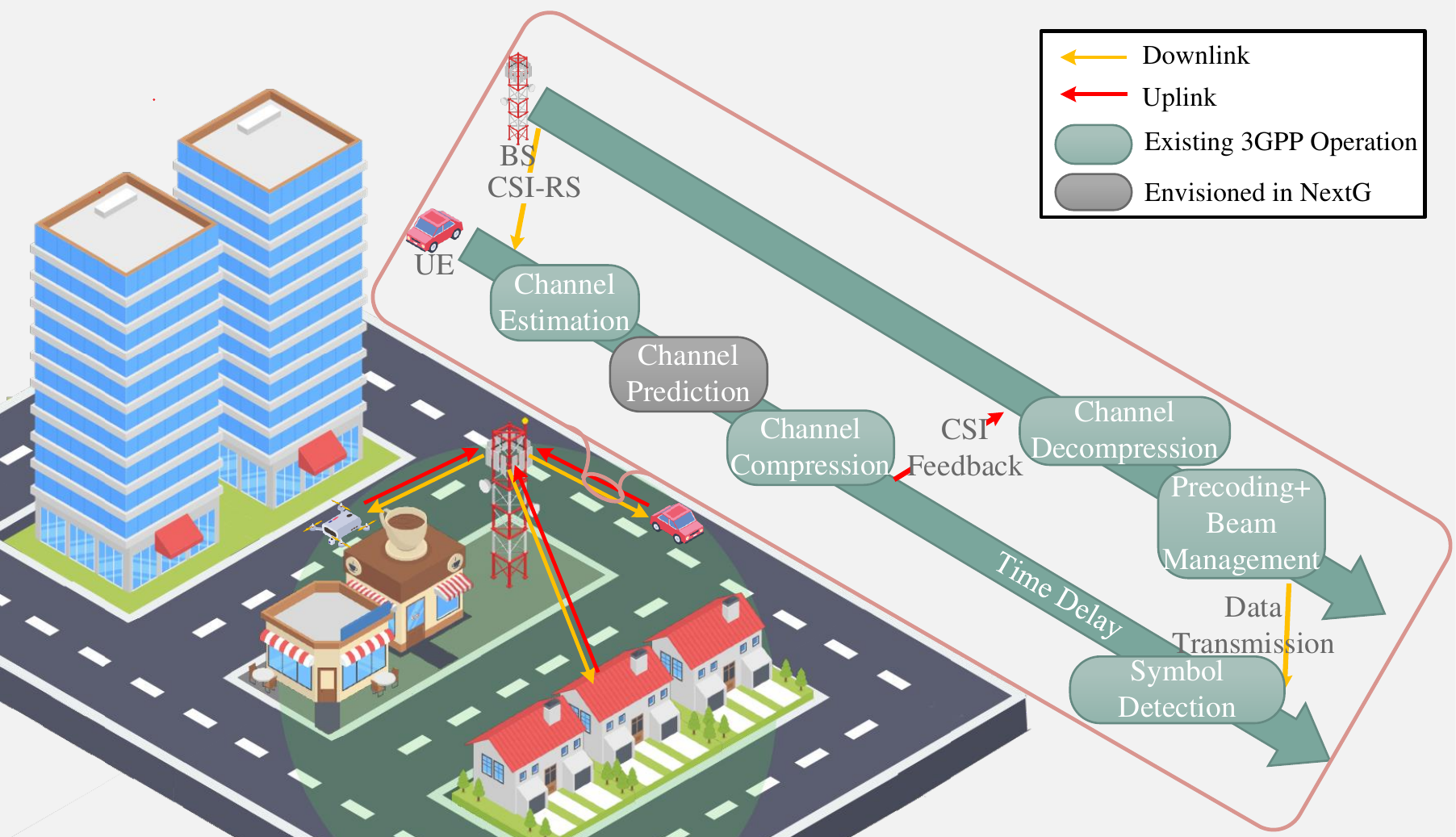}
    \caption{Downlink CSI acquisition and data transmission in an FDD system.}
    \label{fig:overview}
\end{figure*}

The development of online real-time learning-based methods at the air interface of NextG networks has two notable common challenges: i) \emph{limited OTA training data}, and ii) \emph{real-time training constraint}.
For MIMO operations where the OTA training data does not directly contain the target (e.g., channel estimation), an additional challenge is iii) \emph{lack of ground truth}.
In the context of online real-time learning, the OTA training data in the form of standard-mandated RS is notably limited.
The requirement of model updates based on the scheduling granularity calls for fast, low-complexity, and efficient procedures for real-time training.
Moreover, in certain MIMO operations, the absence of explicit ground truth during the online stage exacerbates the complexity of the problem.

In this section, we narrow our focus to MIMO operations within the downlink channel state information (CSI) acquisition and subsequent downlink data transmission processes in a frequency division duplexing (FDD) system. 
We explore the necessity, challenges, and existing works related to online real-time learning-based approaches for these specific operations. 
In the remainder of this article, AI/ML-enabled methods are referred to as ``learning-based approaches''.

\subsection{Overview}
Knowledge of the wireless channel between the transmitter and the receiver (in both downlink and uplink) is critical for ensuring overall end-to-end QoS.
Considering FDD systems as an example as depicted in Fig.~\ref{fig:overview}, downlink CSI acquisition involves a UE estimating the CSI with the CSI RS (CSI-RS).
The estimated CSI is compressed and sent as CSI feedback by the UE to the BS through the uplink control channel. 
Upon reception of the compressed CSI feedback, the BS performs decompression, precoding, and beam management for efficient downlink data transmission. 
The transmitted data is recovered through symbol detection at the UE.
However, a time delay in the CSI acquisition process can lead to outdated feedback, a phenomenon termed ``channel aging'', which can negatively impact the downlink QoS.
While the CSI acquisition process can be simplified in the time-division duplex (TDD) system by leveraging channel reciprocity to avoid CSI feedback, the imperfect channel reciprocity due to high user mobility or hardware impairments may also cause outdated CSI~\cite{8801923, 9804217}.
Therefore, channel prediction has been envisioned as a crucial step to address the channel aging issue in NextG networks.

\subsection{Online Real-Time Channel Estimation}



The accuracy of channel estimation is pivotal for subsequent operations, such as CSI feedback, transmit precoding, beam management, and multi-user scheduling. 
These operations collectively determine whether the link-level and system-level QoS metrics are satisfactorily met.
Conventional channel estimation approaches, such as the least squares (LS) and the linear minimum mean square error (LMMSE) methods, either suffer from inferior estimation performance or require accurate channel statistics.
Due to the limitations of conventional methods, learning-based approaches have started to receive significant attention.
To enhance generalization capability, learning-based approaches with online and real-time capabilities are preferable, particularly in highly dynamic channel environments.

Applying online real-time learning to the channel estimation task has certain key challenges.
Firstly, explicit ground truth in terms of perfect channel knowledge is not available. 
The OTA training data for channel estimation refers to the transmitted and received CSI-RS, instead of the actual channel.
Secondly, the OTA training data in the form of CSI-RS is extremely limited due to the overhead constraint, as mandated by 3GPP standards.
Finally, updating the AI/ML model within the scheduling granularity necessitates extremely low complexity and efficient training procedures.

Recent efforts have aimed at achieving online channel estimation solely using the OTA training data, eliminating the need for perfect channel knowledge. 
Specifically, existing approaches showcase the possibility of employing innovative loss functions~\cite{XZheng2021} and utilizing deep reinforcement learning (DRL)~\cite{MOh2021} for online channel estimation without knowing the perfect channel.
However, these methods fall short of meeting real-time learning requirements, necessitating an extensive number of subframes for convergence. 
To address this, we introduce StructNet-CE in~\cite{li2023learning}, which is an innovative online real-time learning approach that leverages the symmetry of modulation constellations. 
This significantly enhances neural network (NN) training efficiency, facilitating online subframe-based channel estimation.
Given the pivotal role of channel estimation, further exploration in this direction is essential in advancing online real-time learning solutions.

\subsection{Online Real-Time Symbol Detection}

Symbol detection is a critical operation within contemporary communication receivers, aimed at recovering the transmitted information symbol from the received signal.
Conventional symbol detection approaches usually rely on explicit system modeling and accurate knowledge of CSI, which may not always be applicable in real-world scenarios.
Therefore, learning-based approaches offer a promising alternative by circumventing these assumptions.
The dynamic nature of channel environments further calls for an online and real-time symbol detection approach with robust generalization ability.

Designing such a symbol detection method can be challenging due to the scarcity of OTA training data and the real-time training constraint.
Different from channel estimation, the OTA training data for symbol detection is available in the form of the demodulation reference signal (DM-RS), which can serve as ground truth labels.
In a multi-user downlink setting with the BS employing rate-splitting multiple access (RSMA), the model-based learning approach in~\cite{Cerna2023} provides a promising solution for online user-specific symbol detection by designing the algorithm based on successive interference cancellation.
However, its performance with 3GPP-compliant OTA training data and real-time training constraints remains to be seen.
Our prior work~\cite{JXu2023} achieves online real-time symbol detection by leveraging reservoir computing (RC) and embedding structural knowledge in its design.
RC is a unique family of recurrent neural networks (RNN) characterized by having only a few trainable weights, while the majority of the weights are randomly initialized and remain fixed, facilitating a fast and efficient training procedure with limited training data.
In our most recent work~\cite{JereMILCOM2023}, we introduce an approach to initialize the untrained weights of RC based on domain knowledge to further improve its performance for online real-time symbol detection.
We envision that RC can play an instrumental role in enabling online real-time learning solutions, especially for the symbol detection task.



\subsection{Channel Prediction}

Channel aging drives the need for channel prediction, especially in scenarios with moderate to high user mobility in MU-MIMO systems.
However, deriving closed-form solutions for the prediction of non-stationary channel processes is challenging using traditional signal processing and optimization methods. 
Learning-based approaches, on the other hand, can present promising alternatives for CSI prediction.
Online real-time learning methods are preferable over offline approaches for their better generalization ability in dynamic wireless channels encountered during practical deployments.

Designing online real-time channel prediction approaches faces challenges similar to those in the channel estimation task. 
However, the hurdle of lack of ground truth becomes more pronounced in channel prediction, when historical channels are adopted as input to a NN for prediction.
Without perfect channel knowledge, the accuracy of estimated historical channels can significantly influence prediction performance.
For instance, the shallow RNN in~\cite{8801923} addresses limited training data by using a single hidden layer but faces performance loss with inaccurately estimated historical channels during online stages. 
In~\cite{9804217}, DRL is employed for joint channel prediction and beamforming, which optimizes the transmission sum rate to implicitly address ground truth channel absence. 
While the above works enable online learning, their extended training times make adhering to real-time training constraints impractical in real-world systems. 
The design of online real-time channel prediction is still an active area to explore.

\subsection{CSI Compression}

CSI compression plays a crucial role in ensuring robust downlink transmission and achieving high CSI reconstruction accuracy in practical wireless systems.
In massive MIMO systems, where the BS is equipped with a large number of antennas, the resulting CSI dimension becomes extensive, leading to substantial feedback overhead. 
Therefore, there is a need to explore learning-based approaches for CSI compression, especially with online real-time characteristics to address the common generalization issue.

Enabling online real-time learning for CSI compression faces challenges similar to those encountered in channel estimation and prediction. 
Besides these challenges, the inherent difficulty in updating the end-to-end NN from UE to BS presents an additional obstacle in achieving online real-time CSI compression. 
Recent efforts explore adapting offline-trained autoencoders online.
For instance, an unsupervised method is introduced in~\cite{9803260} by adding an extra encoder after the pre-trained decoder at the BS to perform online adaptation. 
While this approach provides a potential solution to address the issue of lack of perfect channel knowledge, the extended processing time for multiple training rounds may be impractical in real-world systems, particularly when CSI feedback occurs every few milliseconds.
The design of NN models for CSI compression that can work in an online real-time learning fashion requires further exploration.

\subsection{Beam Management}



A beam management (BM) framework has been introduced by 3GPP that relies on two sets of downlink beams labeled `Set A' and `Set B', as defined in 3GPP TR 38.843.
This framework utilizes the measurements of reference signal received power (RSRP) resulting from `Set B' (drawn from a low-resolution codebook) to predict the best beam from `Set A' (drawn from a high-resolution codebook) for downlink data transmission.
This is a challenging sparse signal recovery problem, with the only ground truth being the limited RSRP measurements corresponding to the low-resolution codebook.
Although iterative compressive sensing (CS)-based optimization methods can be utilized, their computational complexity becomes prohibitive in massive MIMO systems.
On the other hand, learning-based methods can potentially provide computationally efficient solutions. 
The vulnerability of BM to user mobility in the mmWave bands due to narrower beam widths requires the NN model weights at the BS updated upon receiving the periodic RS (e.g., CSI-RS).
Therefore, learning-based approaches with online and real-time characteristics are necessary to tackle high user mobility for BM, especially in mmWave systems.


The common challenges of online and real-time learning are also applicable to BM.
Namely, the training data in the form of periodic RSRP reports from the served UEs is extremely scarce, while explicit ground truth labels (drawn from the high-resolution codebook) are not available. 
Additionally, the real-time constraint of executing the beamforming decision is highly stringent.
Therefore, for robust adaptation to user mobility, the online learning approach must be able to use extremely limited OTA training data and have a short update time to facilitate standards-compliant implementations.

As an example, a DRL-based beam training framework that mitigates the challenge of lack of explicit ground truth data and misaligned scenarios is introduced in~\cite{Zhang2021_BeamMgmt}. 
Although such DRL-based approaches can meet online learning requirements, they are typically not amenable to real-time training due to long convergence times, potentially limiting their applicability in highly non-stationary channel scenarios caused by rapid user mobility. 
Thus, developing online learning-aided methods for BM with low computational complexity remains an active area of investigation.



\section{Design Insights for Online Real-Time Learning}


\subsection{Design Insights and Guidelines}

\begin{figure}[t] 
    \centering
    \includegraphics[width=1\linewidth]{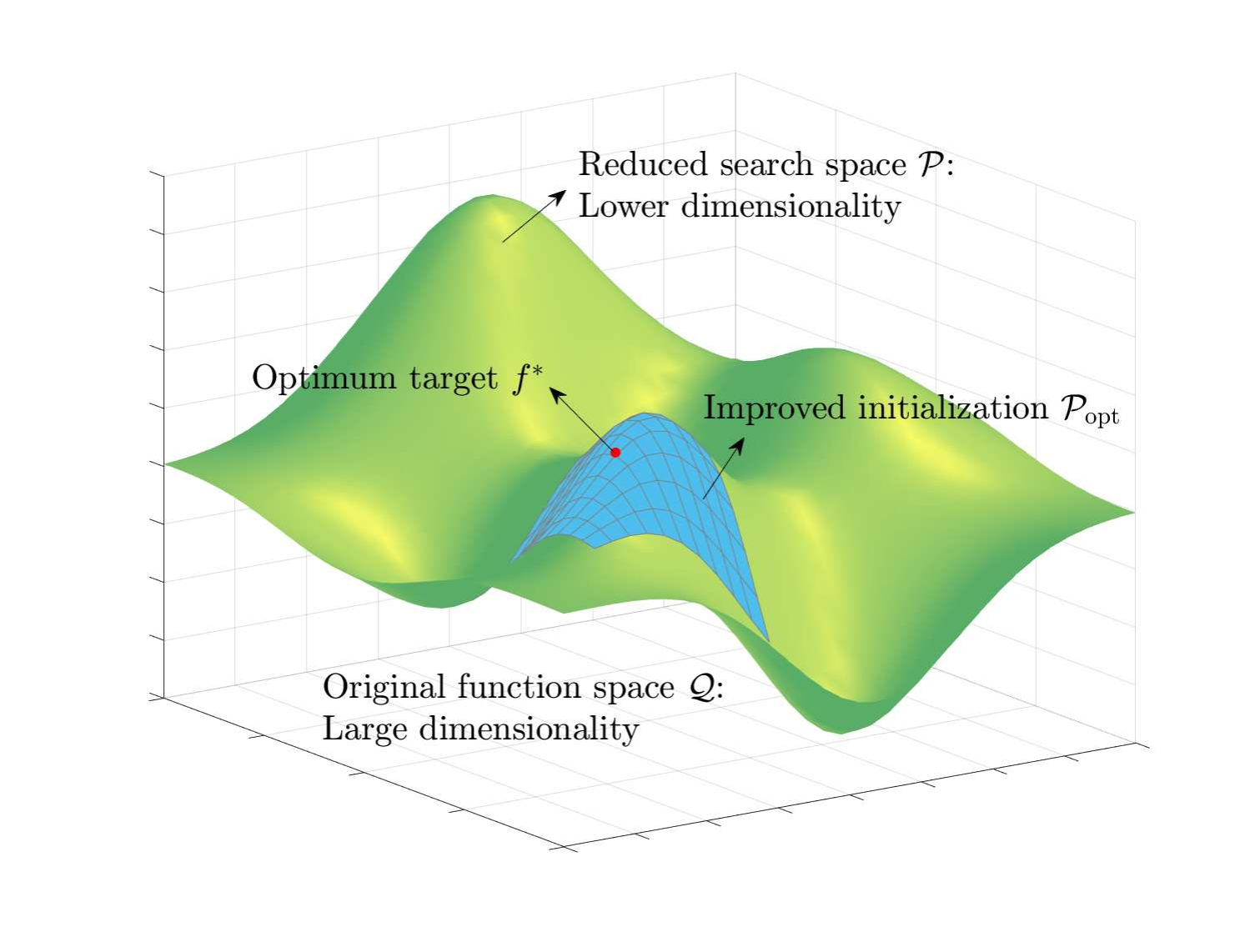}
    \caption{Exploiting domain knowledge to aid online learning.}
    \label{fig:domain_knowledge}
\end{figure}
In light of the preceding challenges, we foresee three possible approaches to achieve online real-time learning:
i) \textbf{Reduce dimensionality of search space:} 
A possible approach is to reduce the dimensionality of the search space based on domain knowledge, ensuring the target is in the search space.
By reducing the dimensionality of the search space, a smaller NN model with fewer trainable parameters can be employed to facilitate online real-time learning in a sample-efficient manner.
This is illustrated in Fig.~\ref{fig:domain_knowledge}, where exploiting domain knowledge permits confining the search space to a lower dimensional function space $\mathcal{P}$ instead of the original function space $\mathcal{Q}$ of much larger dimensionality.
As an example, the symmetric structure of modulation constellations can be exploited to transform the multi-class classification problem of symbol detection into tasks that can be solved by a single binary classifier, as shown in our previous work~\cite{JXu2023}. 
ii) \textbf{Improve initialization:}
A supplementary strategy is to expedite online learning by initializing NN weights based on domain knowledge. 
This ensures that the initial starting point for model training in the reduced search space is drawn from a neighborhood that is much closer to the target than an arbitrary random model initialization in $\mathcal{P}$, thus enabling faster convergence to the optimum target.
This is conceptually depicted in Fig.~\ref{fig:domain_knowledge}, where domain knowledge is exploited further to initialize training within the neighborhood $\mathcal{P}_{\mathrm{opt}}$ around the target model $f^{*}$.
For example, for the task of online real-time symbol detection, a novel procedure was introduced in our most recent work~\cite{JereMILCOM2023} to initialize the untrained weights of an RC-based symbol detector based on domain knowledge, specifically knowledge of the channel statistics, thereby resulting in a vast improvement in performance over an RC-based detector with randomly initialized untrained weights, i.e., without domain knowledge utilization.
iii) \textbf{Select proper learning objectives:}
Under circumstances where the ground truth label is not available, the learning objectives can be designed based on domain knowledge to enable the NN to estimate the target without knowing the ground truth.
Specifically, instead of designing the objective function to explicitly estimate the target, the learning objective can be designed to aim at another task that is related to the target based on domain knowledge, making the target estimation an implicit optimization goal.
For instance, as demonstrated in our prior work~\cite{li2023learning}, the channel can be implicitly estimated by optimizing over the symbol detection task and incorporating the symmetry of modulation constellations into the NN architecture. 
As another example, the channel prediction problem can be solved by maximizing the received downlink sum rate through DRL~\cite{9804217}, instead of explicitly minimizing the prediction error.

\subsection{Case Study: Online Real-Time MIMO Channel Estimation}
\label{sec:online_channel_estimation}





\begin{figure}[t] 
    \centering
    \includegraphics[width=\linewidth] {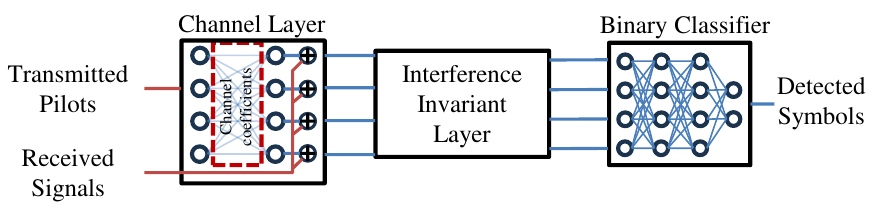}
    \caption{Structure of StructNet-CE.}
    \label{fig:StructNet-CE}
\end{figure}


Given the aforementioned insights, we present an online real-time MIMO-OFDM channel estimation approach named StructNet-CE~\cite{li2023learning}.
StructNet-CE leverages the symmetric structure of the QAM constellation to reduce the dimensionality of the search space, which equips it with a fast and efficient learning procedure.
With the reduced search space dimensionality and improved sample efficiency, StructNet-CE can learn with only the OTA training data and conduct channel estimation on a subframe basis, achieving the online and real-time characteristics.
Furthermore, StructNet-CE is optimized over the symbol detection task while achieving channel estimation as an implicit optimization goal, addressing the challenge of lacking perfect channel knowledge during the online stage.
The training data utilized for learning StructNet-CE is the transmitted and received OTA RS.


\begin{table}[tb]
\centering
\caption{Experimental settings}
\label{tab:setting}
\resizebox{1\columnwidth}{!}{
\begin{tabular}{|c|c|c|}
\hline
\textbf{Item}                                 & \textbf{Offline}                 & \textbf{Online}                 \\ \hline
Receive antenna number                & 2                       & 2                      \\ \hline
Transmit antenna number                & 1                       & 2                      \\ \hline
Channel scenario                        & Rural macro-cell non-line-of-sight (NLOS) & Urban macro-cell NLOS \\ \hline
User speed                           & 50 km/h                & 5 km/h\\ \hline
Subcarrier number                    & \multicolumn{2}{c|}{1024}                         \\ \hline
OFDM symbols per subframe  & \multicolumn{2}{c|}{14}                           \\ \hline
Data symbols per subframe  & \multicolumn{2}{c|}{10}                           \\ \hline
Pilot symbols per subframe & \multicolumn{2}{c|}{4}                            \\ \hline
\end{tabular}
}
\end{table}

As illustrated in Fig.~\ref{fig:StructNet-CE}, StructNet-CE consists of a channel layer for channel estimation, an interference invariant layer to capture inter-stream interference in single-user MIMO, and a binary classifier for symbol detection.
This architecture leverages the symmetry of the modulation constellation, allowing the transformation of a multi-class classification problem into a series of binary detection tasks. 
The channel layer and the binary classifier are designed to facilitate this conversion. 
Moreover, the detection of symbols in the target data stream should remain unaffected by the symbols transmitted in interference data streams.
The interference invariant layer is designed to capture this property.
More details about the design are provided in~\cite{li2023learning}.
The input to StructNet-CE is the transmitted pilot along with the received signal, and the output is the detected symbol.
By embedding the constellation symmetry, StructNet-CE forces the weights of the channel layer to be aligned with the MIMO channel coefficients through backpropagation.
This allows for the direct acquisition of the estimated channel coefficients from the channel layer weights without perfect channel knowledge.
To demonstrate the advantages of online real-time learning, we conduct an experiment where a large NN model, specifically ReEsNet~\cite{LLi2020}, is considered for channel estimation, with both channel scenarios and system configurations used for offline training and online testing being misaligned.
Subsequently, its performance is compared with StructNet-CE, which is trained completely online.
Note that to the best of our knowledge, online adaptation of the offline-trained ReEsNet without the ground truth channel being available is still a challenging problem, and therefore is not considered here.
The simulation settings during the offline and online stages are summarized in Tab.~\ref{tab:setting}.
More detailed settings are provided in~\cite{li2023learning}.
The channel realizations are generated by QuaDRiGa using the 3GPP-3D channel model specified in 3GPP TR 36.873.
The adopted pilot pattern complies with the 3GPP standard, which follows Tab. 7.4.1.1.2-4 in 3GPP TS 38.211.
Orthogonal pilots are employed for conventional methods, while non-orthogonal pilots are exploited for learning-based approaches, following the discussion of pilot choice in~\cite{JXu2023}.
Note that when adopting ReEsNet trained offline under the $1\times 2$ SIMO-OFDM system to the $2\times 2$ MIMO-OFDM online scenario, the $2 \times 2$ MIMO-OFDM channel is estimated via two $1\times 2$ SIMO-OFDM systems without considering the inter-stream interference.

\begin{figure}[t] 
    \centering
    \includegraphics[width=1\linewidth] {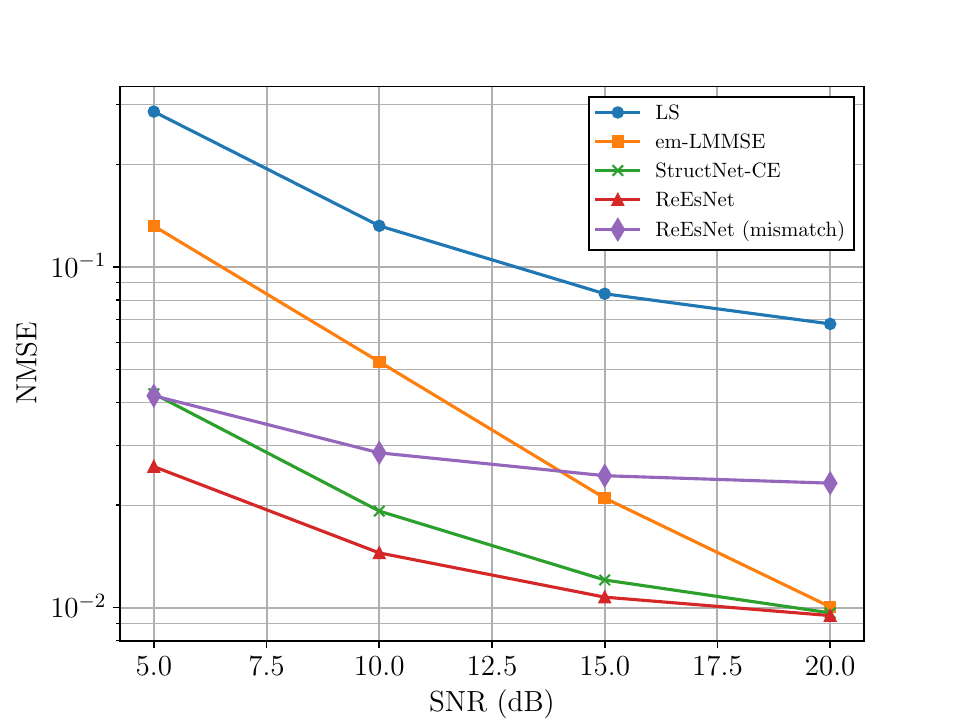}
    \caption{NMSE of channel estimation methods. 
    }
    \label{fig:MSE}
\end{figure}

Fig.~\ref{fig:MSE} compares the normalized mean square error (NMSE) of different approaches under various signal-to-noise ratios (SNRs).
The SNR is defined as the ratio of the received signal power to the noise power.
Compared with ReEsNet without such misalignment (labeled ``ReEsNet"), ReEsNet with mismatched offline-online channel scenarios and system configurations, denoted as ``ReEsNet (mismatch)", exhibits performance degradation due to the ``uncertainty of generalization" issue. 
While ReEsNet without any misalignment outperforms StructNet-CE, such performance is an unachievable error lower bound for ReEsNet.
This is because it is extremely difficult in practice to ensure the online deployment scenario matches exactly with the scenario encountered during offline training, due to the dynamic channel environment.

In contrast, the purely online and real-time trained StructNet-CE outperforms ReEsNet with mismatches as well as conventional approaches, such as LS and empirical LMMSE (labeled ``em-LMMSE"), while utilizing only the OTA pilots for training and adhering to real-time constraints.
More importantly, in the high SNR regime, StructNet-CE can reach the practically unachievable error lower bound of ReEsNet, demonstrating the effectiveness of StructNet-CE in online real-time learning.
In~\cite{li2023learning}, the CPU run time is furnished to demonstrate the real-time properties of StructNet-CE.

In summary, StructNet-CE overcomes the bottleneck of poor generalization seen with purely offline training by effectively incorporating domain knowledge into its design. 
With purely online and real-time learning, StructNet-CE outperforms conventional approaches and achieves the error lower bound that cannot be obtained by ReEsNet in the high SNR regime.

\section{Conclusion and Outlook}

The futuristic use cases envisioned in NextG will bring about the need for sophisticated MIMO techniques, where conventional model-based solutions may fall short.
AI/ML techniques can come to the rescue to realize such applications.
In this article, a pivotal role of AI/ML-based solutions is envisioned, particularly of online real-time learning methods to implement critical MIMO operations at the air interface.
The challenges involved in designing online real-time AI/ML-based methods for these tasks are outlined, and supported with crucial insights that can serve as guiding principles in constructing such approaches.
Building on these insights, an online real-time AI/ML-based solution is introduced for channel estimation.
Looking ahead, we believe that effective utilization of domain knowledge in the design of the structure of AI/ML models, which remains an exploratory endeavor at present, will be central in making both online real-time learning and general hybrid learning approaches a reality at the NextG air interface.






\ifCLASSOPTIONcaptionsoff
  \newpage
\fi



\bibliographystyle{IEEEtran}

\bibliography{IEEEabrv,ref.bib}
%

%


\begin{IEEEbiographynophoto}{Jiarui Xu}
is a Ph.D. student at Virginia Tech, USA.
\end{IEEEbiographynophoto}
\begin{IEEEbiographynophoto}{Shashank Jere}
is a Ph.D. student at Virginia Tech, USA.
\end{IEEEbiographynophoto}
\begin{IEEEbiographynophoto}{Yifei Song}
is a Ph.D. student at Virginia Tech, USA.
\end{IEEEbiographynophoto}
\begin{IEEEbiographynophoto}{Yi-Hung Kao}
is a Ph.D. student at Virginia Tech, USA.
\end{IEEEbiographynophoto}
\begin{IEEEbiographynophoto}{Lizhong Zheng}
is a Professor in Electrical Engineering and Computer Science at Massachusetts Institute of Technology, USA.
\end{IEEEbiographynophoto}
\begin{IEEEbiographynophoto}{Lingjia Liu}
is a Professor \& Bradley Senior Faculty Fellow in Electrical and Computer Engineering at Virginia Tech, USA.
\end{IEEEbiographynophoto}





\end{document}